\documentclass[a4paper,12pt]{article}
\usepackage{graphicx}

\newcommand{\mc}{m_{\rm c}}
\newcommand{\Tc}{T_{\rm c}}
\renewcommand{\vec}[1]{{\stackrel{\rightarrow}{#1}}}

\begin{document}

\begin{titlepage}
\title{Chiral quark model with infrared cut-off for the
description of meson properties in hot matter}

\author{
D. Blaschke, G. Burau\\[2.5mm]
 Fachbereich Physik, Universit\"at Rostock,\\
D-18051 Rostock, Germany\\[6mm] 
 M. K. Volkov, and V. L. Yudichev\\[2.5mm]
 Bogoliubov Laboratory of Theoretical Physics,\\
Joint Institute for Nuclear Research,\\
141980 Dubna, Russian Federation
}
\end{titlepage}
\maketitle

\begin{abstract}
A simple chiral quark model of the Nambu--Jona-Lasinio (NJL) type with a
quark confinement mechanism is constructed for the description of the
light meson sector of QCD at finite temperature.
Unphysical quark production  thresholds in the NJL model are excluded by an
infrared cut-off in the momentum integration within quark loop diagrams.
This chiral quark model satisfies the low energy theorems.
Using the vacuum masses and decay widths of  $\pi$- and $\rho$-mesons
for fixing the model parameters, the properties of the $\sigma$- meson
are derived.
Within the Matsubara formalism, the model is systematically extended to
finite temperatures where chiral symmetry restoration due to a dropping
constituent quark mass entails a vanishing of the infrared cut-off
(deconfinement) at the pion Mott temperature $T_{c}=186$ MeV.
Besides of the pion mass and weak decay constant, the masses, coupling
constants and decay widths of  $\sigma$- and $\rho$-mesons in hot matter
are investigated.
The quark-antiquark decay channel of the light mesons is opened for $T>T_c$
only and becomes particularly strong for the $\rho$- meson.
The two-pion decay channel below $T_c$ has almost constant width for the
$\rho$- meson up to $T_c$, but for the $\sigma$-meson it closes below $T_c$
such that a scalar meson state with vanishing width is obtained as a precursor
of the chiral/deconfinement transition.
\end{abstract}


\clearpage
\section{Introduction}
One of the most interesting phenomena predicted by QCD
for a hot and dense matter is
the existence of the quark-gluon plasma (QGP) phase \cite{QGP2}, where
hadrons do not exist as bound states, and the strongly interacting
matter should be described in terms of QCD fundamental fields:
quarks and gluons.
A number of modern experiments have been already
directed on the search of
signals that can be interpreted as
evidence for the existence of the QGP, and still
new projects are upcoming. The most straightforward experiment, where
QGP is hoped to be found is a collision of heavy ions with
ultrarelativistic energies, where hadrons form so hot and dense
matter that the conditions necessary for QGP creation are fulfilled.
However, one cannot tell certainly
if QGP was seen or not without laying a proper theoretical ground
for such an experiment. Actually, one needs to know what is to be
expected in order to unambiguously detect the QGP is formed.
Taking into account that a direct modelling of QGP from QCD is
not available at present, one needs an appropriate quark model which
would reflect the most important features of strong interactions,
on the one hand, and would be tractable, on the other hand.
Among the various approaches, one can select the Nambu--Jona-Lasinio
(NJL) model.

The Nambu--Jona-Lasinio  model is a convenient
semi\-phenomenological quark model for the description of the
low energy meson physics \cite{eguchi,ev83,volk86,ebert86,vogl}.
Within this model the mechanism of spontaneous breaking of chiral
symmetry (SBCS) is realized in a simple and transparent way, and
the low energy theorems are fulfilled.

However, the ordinary NJL model fails to prevent had\-rons from
decaying into free quarks, which makes the realistic description of
had\-ron properties on their mass shell questionable.
The solution of this problem seems to be a very difficult task,
and different methods have been proposed for this purpose
\cite{efimov,grossmilana,celenza95,rw94,bbkr,erf96}.
In our previous papers \cite{volk98,IRconfT00,IRconfTYF},
a chiral quark model  was
suggested   where unphysical quark-antiquark thresholds were
eliminated by means of  an infrared (IR) cut-off.
As a result, the pole of the integrand in a quark loop integral
turned out to be outside the integration interval, so that no imaginary part
occurs.

\begin{figure}[h]
\begin{center}
\includegraphics[scale=1]{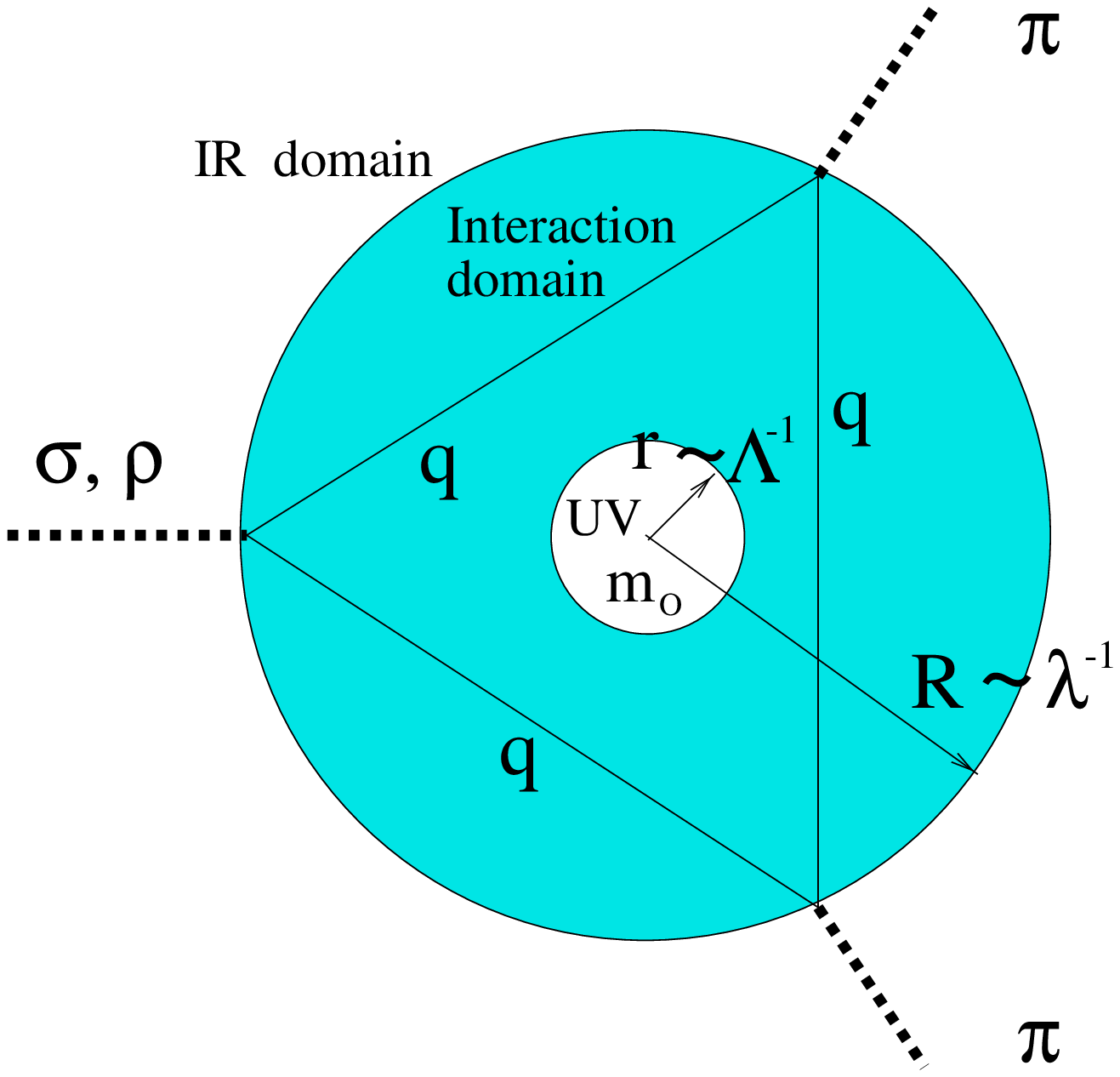}
\end{center}
\caption{}
\label{domains}
\end{figure}

This method of modeling the phenomenon of confinement
is based on the idea of combining the NJL
\cite{eguchi,ev83,volk86,ebert86,vogl}  and bag models
\cite{bag,bag1,bag2}.
 Together with the ultraviolet (UV) cut-off, which is necessary
to eliminate the UV divergence, we introduce the IR
cut-off and thereby divide the momentum space into
three domains. In Fig.~\ref{domains} these domains are represented
schematically in the coordinate space.

The first domain corresponds to short distances (large momenta),
where quarks are not confined and the chiral symmetry is not
spontaneously broken. This domain is excluded by the UV cut-off $\Lambda$.

In the second domain, SBCS takes place, and the quark condensate
appears, which leads to the replacement of the current quarks by constituent
ones.  According to the idea of the bag-model, it is necessary to assume
that quarks are not allowed to propagate over distances exceeding the
dimensions of a bag.
In the language of quark-loop integrals, this leads to a low momenta,
infrared (IR),
cut-off $\lambda$ that provides the confinement of quarks.

In  \cite{volk98,IRconfT00,IRconfTYF}, the IR cut-off
$\lambda$ was chosen to be proportional to the constituent quark mass $m$:
$\lambda=c\; m$, where the arbitrary parameter $c$ was the same for all
mesons.  At $c\sim 2$, the nonphysical quark-antiquark thresholds were
absent for the $\pi$-, $\sigma$-, and $\rho$-mesons in the vacuum.  However,
in hot matter, these thresholds appeared at certain temperatures
which, on the one hand, did not coincide with the chiral symmetry
restoration (CSR) temperature $T_c$ and, on the other, were different for
various sorts of mesons.

In the present work, we suggest a new version of the model with an IR
cut-off which differs from that given in
\cite{volk98,IRconfT00,IRconfTYF} in two main points.  First,
we suggest that the IR cut-off is different
for various mesons and increases with the mass of the
corresponding meson.  Second, the deconfinement phase transition
occurs at a certain temperature $T_{\rm c}$, unique for all mesons, which
coincides with the CSR temperature.
In order to define this critical temperature uniquely also away from the
chiral limit, we will identify it with the pion Mott temperature.
Above this temperature the pion (and also the other mesons) are no longer
bound states but resonances in the $q\bar q$ continuum instead.

We use our model to investigate the behavior of the constituent quark
mass, the pion weak decay constant $F_\pi$, and the masses and
coupling constants of the $\pi$-, $\sigma$- and $\rho$-mesons within
the temperature interval up to the CSR and deconfinement phase
transition.  Very interesting effects at high temperature are revealed
for the $\sigma$ particle.  The $\sigma$-meson, while having a noticeable
width for the decay into two pions in the vacuum, can appear as a very
narrow resonance in some processes occurring in hot and dense matter.
This may lead to  a significant enhancement of the dilepton and
$\gamma\gamma$ production in the processes $\pi\pi\to e^+e^-$ and
$\pi\pi\to\gamma\gamma$ near the phase transition
\cite{PhotonExcess,Kuraev98}.
Detailed calculations of the effect of excess low-energy photon
production in the process $\pi\pi\to\gamma\gamma$ have been performed
in \cite{Kuraev98} (see also \cite{Hatsuda94}).  These effects can be
understood as precursors of the phase transition between hadron matter
and the QGP to be investigated in heavy-ion collisions
performed e.g. at CERN SPS, at the new Brookhaven relativistic heavy
ion collider (RHIC) \cite{QM97} or at future facilities dedicated to the QGP
search.
Note that in \cite{PhotonExcess,Kuraev98} a simple version of the NJL
model was used, where the confinement of quarks had not been taken
into account. Insofar as the confinement is important for the
study of the phase transition of hadron matter to QGP, we then
suggest here a new version of the model, with  quark confinement.

Our paper is organized as follows. In Sect.~2, we give the effective
chiral quark Lagrangian and the gap equation describing SBCS.  Here,
the pion mass formula in the IR cut-off scheme is obtained, and it is
shown that the pion is a Goldstone boson in the chiral limit.  In
Sect.~3, the scalar meson ($\sigma$) is considered, and it is
demonstrated that the quark loop with two $\sigma$-meson legs does not
have an imaginary part when the IR cut-off $\lambda$ is used. The model
parameters are fitted in Sect.~4.   In Sect.~5,
the $\sigma$-meson mass and the $\sigma\to 2\pi$ decay width are
estimated.   In Sect.~6, we introduce a finite temperature
generalization of the model and analyze  results.
In the last section, we discuss the obtained results and give a short outlook
to  further development  of this model.

\section{SU(2)$\times$SU(2) Lagrangian, the gap equation and the pion mass
formula}

Let us consider an SU(2)$\times$SU(2) NJL model defined by the Lagrangian
\begin{eqnarray}
{\mathcal L}_q&=& {\bar q}(i{\partial\hspace{-.5em}/\hspace{.15em}}-m^0)q
+ {G_1\over 2} \left[({\bar q} q)^2 +({\bar q}i{\gamma}_5
\vec{\tau} q)^2\right]\nonumber\\
&&-\frac{G_2}{2}\left[ (\bar q\gamma_\mu\vec\tau q)^2+
(\bar q\gamma_5\gamma_\mu\vec\tau q)^2\right]
\label{Lq},
\end{eqnarray}
where $q$ and $\bar q$ are quark fields, and $G_1$ and $G_2$ are
the constants that describe the interaction of quarks in the
scalar (pseudoscalar) and vector (axial-vector) channels.
The notation $\vec{\tau}$ is used for Pauli matrices, represented
here as components of an isovector. This Lagrangian is chirally
symmetric except for the term containing the current quark mass $m^0$.

As usual, one applies a bosonization procedure \cite{ev83,volk86} to the
quark Lagrangian and  obtains its equivalent representation in terms of the
scalar ($\sigma$),
 pseudoscalar ($\vec{\pi}$), vector ($\vec{\rho}_\mu$) and
axial-vector ($\vec{a}_{1\,\mu}$) meson fields
\begin{eqnarray}
\label{lmes}
&&{\mathcal L}_{\rm mes} = -\frac{\tilde{\sigma}^2+\vec{\pi}^2}{2 G_1}
+\frac{\mathop{\vec \rho_\mu }^2+\mathop{\vec{ a}}_{1\,\mu}^2}{2 G_2}
\nonumber\\
&&\quad- i {\rm Tr}\! \ln\left\{1\!+\!\frac{1}{i\partial\hspace{-.5em}/\hspace{.15em} -m}
[\sigma\! +\! i \gamma_5\vec{\tau}\vec{\pi}\!+\!
\vec\tau{\hat{\vec \rho}}\!+\!\gamma_5\vec\tau{\hat{\vec a}}_{1}
]\right\}.
\label{lag2}
\end{eqnarray}
Here, the scalar fields $\sigma$ and $\tilde{\sigma}$  are  connected by the
relation
\begin{eqnarray}
-m^0 + \tilde{\sigma} &=& -m + \sigma ~.
\end{eqnarray}
The vacuum expectation value of $\tilde\sigma$
is non-zero after SBCS and new notation for the fluctuating part $\sigma$
of the scalar field with zero vacuum expectation value,
$\langle \sigma \rangle_0= 0$, is introduced here.
The vector $\hat{\vec\rho}$
and axial-vector $\hat{\vec a}_1$ fields are members of the Dirac
algebra:  $\hat{\vec\rho}={\vec\rho}_\mu\gamma^\mu$,
$\hat{\vec a}_1={\vec a}_{1\,\mu}\gamma^\mu$.
Then, from the condition
\begin{eqnarray}
\frac{\delta {\mathcal L}_{\rm mes}}{\delta \sigma}\Bigg|_{\sigma=0,
\vec{\pi}=0} =0~~,
\end{eqnarray}
one obtains the gap equation
\begin{eqnarray}
&&m^0=m[1-8 G_1 I_1^{\Lambda}(m)]
=m+2G_1\langle\bar qq\rangle_0~,
\label{gap}
\end{eqnarray}
where $\langle \bar q q \rangle_0$ is the quark condensate.
The quantity $I_1^{\Lambda}(m)$ is obtained from the
quadratically divergent integral $I_1^{\infty}(m)$ by regularization
with the UV cut-off $\Lambda$,
\begin{eqnarray}
&&I_1^{\Lambda}(m)=-i\frac{N_c}{(2\pi)^4}\int
\frac{d^4 k}{m^2 - k^2 -i\varepsilon} \theta(\Lambda^2-|k_\perp|^2)
\nonumber\\
&&\quad=\frac{N_c}{(2\pi)^2}\int^\Lambda d\/{\rm k}
\frac{\rm k^2}{E_{\rm k}}\nonumber\\
&&\quad=\frac{N_c}{8 \pi^2}\left[\Lambda\sqrt{\Lambda^2+m^2}-{\rm ln}\left(\frac{\Lambda+\sqrt{\Lambda^2+m^2}}{m}\right)\right],
\end{eqnarray}
where $E_{\rm k}=\sqrt{{\rm k}^2+m^2}$,
$k_\perp$ is  the 4-momentum of a quark,
transverse to an arbitrary momentum $P$ ($P^2\not=0$)
(see \cite{NPA95,Volkov96})
\begin{equation}
{k_\perp}_\mu=k_\mu-P_\mu \frac{P\cdot k}{P^2}~,
\end{equation}
so that for $P=(M,0,0,0)$  one has $k_\perp=(0,{\rm k_1,k_2,k_3})$;
$N_c$ is the number of colors.
For all applications of the model,  we  do not introduce
an IR cut-off in $I_1^{\Lambda}(m)$.

\begin{figure}[h]
\begin{center}
\includegraphics[scale=1]{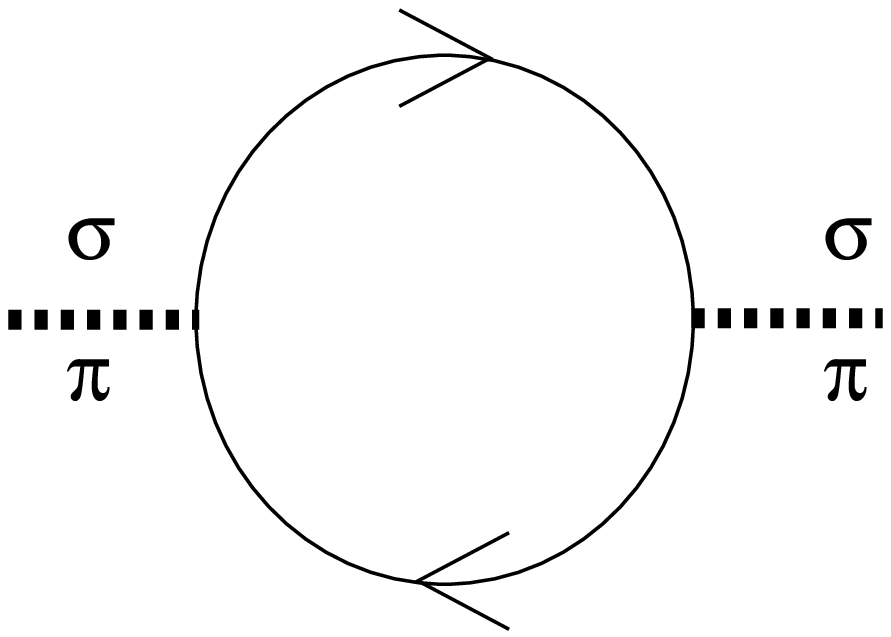}
\end{center}
\caption{}
\label{qloop}
\end{figure}

Now let us consider the free part of  Lagrangian (\ref{lmes}) for pion
fields in the quark one-loop approximation (see Fig.~\ref{qloop})
\footnote{The
expression enclosed in parentheses can be
written in the form $1/G_1+\Pi_\pi(p)$,
where $\Pi_\pi(p)$ is the pion polarization operator.}
\begin{eqnarray}
{\mathcal L}^{(2)}_\pi=-\frac{\vec{\pi}^2}{2}\left\{\frac{1}{
G_1}-8I_1^{\Lambda}(m)
-4 P^2 I_2^{(\lambda_{P}\,\Lambda)}(P,m)\right\}~,
\label{Lpi2}
\end{eqnarray}
where $I_2^{(\lambda_{P},\Lambda)}(P,m)$ is obtained from the
logarithmically divergent integral $I_2^{(0,\infty)}(P,m)$
by applying IR- and UV- cut-offs
\begin{eqnarray}
&&I_2^{(\lambda_P,\Lambda)}(P,m)=\nonumber\\
&&\quad=-i\frac{N_c}{(2\pi)^4}\int\frac{ \theta(\Lambda^2-|k_\perp|^2)
\theta(|k_\perp|^2-\lambda_P^2) d^4 k}{(m^2-k^2-i\varepsilon)
(m^2-(k-P)^2 -i\varepsilon)}\nonumber\\
&&\quad=\frac{N_c}{2 \pi^2}\int_{\lambda_P}^\Lambda
d\/{\rm k}\frac{{\rm k}^2}{E_{\rm k}(4E_{\rm k}^2-P^2)}~.
\label{i2}
\end{eqnarray}
\noindent
Here, $P$ is the momentum of a bound $\bar qq$ state (meson) and
$\lambda_P$ is an infrared cut-off introduced in order to remove unphysical
quark-antiquark production thresholds, see Sect. \ref{sec:sigma}.
The last integration in (\ref{i2}) is done in the rest frame of
a meson $({\bf P}=0)$.
The integral $I_2^{(\lambda_{P},\Lambda)}(P,m)$ thereby depends on a
Lorentz invariant, the meson mass $M$. Further, we prefer to consider
$I_2^{(\lambda_P,\Lambda)}(P,m)$ as a function of $M$:
\begin{equation}
I_2^{(\lambda_M,\Lambda)}(M,m)\equiv\left.I_2^{(\lambda_P,\Lambda)}(P,m)
\right|_{P=(M,0,0,0)}.
\end{equation}

To express (\ref{Lpi2}) through physical fields,
we renormalize the pion
\begin{eqnarray}
\vec{\pi}&=& g_\pi(M_\pi)\vec{\pi}^r, \\
~g_\pi(M_\pi)&=&[4I_2^{(\lambda_{M_\pi},\Lambda)}(M_\pi,m)]^{-1/2}.
\end{eqnarray}

For the pion, there is also an additional renormalization
factor $\sqrt{Z}$ appearing after we take
 into account  $\pi-a_1$ transitions
\cite{volk86}:
\begin{equation}
\bar g_{\pi}=g_{\pi}\sqrt{Z},\qquad
Z^{-1}=1-\frac{6m^2}{M_{a_1}^2},  \label{gpi}
\end{equation}
where  $M_{a_1}=1230$ MeV is the mass of the $a_1$-meson.
Thus, we obtain the following expression for the pion mass:
\begin{equation}
M_\pi^2=\bar g_\pi^2(M_\pi)\left[\frac{1}{ G_1}-8
I_1^{\Lambda}(m)\right].
\label{pionmass}
\end{equation}
It can be given the form of the Gell-Mann--Oakes--Renner relation
\begin{equation}
M_\pi^2 \approx -2\frac{m^0\langle\bar qq\rangle_0}{F_\pi^2},
\label{mpi}
\end{equation}
where the Goldberger-Treiman relation (see (\ref{gold}) below) and the gap
equation (\ref{gap}) have been used.
One can see that this pion mass formula is in accordance with the Goldstone
theorem since, for $m^0=0$, the pion mass  vanishes  and  the pion
becomes a Goldstone boson.

\section{The $\sigma$-meson and IR confinement}
\label{sec:sigma}

The free part of Lagrangian (\ref{lag2}) for the $\sigma$-meson
in the one-loop approximation (see Fig.~\ref{qloop}) has the following form
\begin{eqnarray}
&&\kern-5mm{\mathcal L}^{(2)}_\sigma
=\!\!-\frac{{\sigma}^2}{2}\!\!\left\{\!\frac{1}{G_1}\!-\!8I_1^{\Lambda}(m)\!
-\!4(P^2\!\!-\!4m^2)\! I_2^{(\lambda_{M_\sigma},\Lambda)}(P,m)\!\right\}\!.
\label{Lsig2}
\end{eqnarray}
After the renormalization of the $\sigma$ field,
\begin{eqnarray}
\sigma&=&g_\sigma(M_\sigma)\sigma^r,\\
g_\sigma(M_\sigma)&=&[4I_2^{(\lambda_{M_\sigma},\Lambda)}(M_\sigma,m)]^{-1/2},
\label{gsigma}
\end{eqnarray}
one obtains the expression for the $\sigma$-meson mass
\begin{eqnarray}
M_\sigma^2&=& g_\sigma^2(M_\sigma)\left[\frac{1}{ G_1}-8
I_1^{\Lambda}(m)\right]+4m^2.
\label{msigma}
\end{eqnarray}
Now, let us consider more carefully the integral
\begin{eqnarray}
I_2^{(\lambda_{M_\sigma},\Lambda)}(M_\sigma,m)=
\frac{N_c}{2\pi^2} \int_{\lambda_{M_\sigma}}^{\Lambda}
\! d\/{\rm k} \frac{{\rm k}^2}{E_{\rm k}(4E_{\rm k}^2-M_\sigma^2)}.
\label{i2sigma}
\end{eqnarray}
If $\lambda_{M_\sigma}=0$, this integral has an imaginary part.
Indeed, the  integrand in (\ref{i2sigma})
is singular when its denominator
is equal to zero:
\begin{equation}
4E_{\rm k}^2-M_\sigma^2=0. \label{id}
\end{equation}
The imaginary part appears when the singularity
(${\rm k}=\frac12\sqrt{M_\sigma^2-4m^2}$)
lies within the integration interval.
Therefore, when one applies the IR cut-off
\begin{eqnarray}
&&\lambda_P=
\left[\mc\,\theta(m-\mc)+m\,\theta(\mc-m)\right]\nonumber\\
&&\quad\times\theta(P^2-4\mc^2)\sqrt{\frac{P^2}{4m_{\rm c}^2}-1},
 \label{cutoff}
\end{eqnarray}
the denominator of  integral (\ref{i2sigma}) has no zero if the
new model parameter $m_{\rm c}$ is smaller then the constituent quark mass.
Thus,  integral (\ref{i2sigma})  is real, and the quark-antiquark production
threshold is absent.
We consider this property as a criterion for the quark confinement.
It is different from that of the absence of real mass poles in the quark
propagator, which is employed within the DSE approach \cite{rw94,bbkr}.
The   parameter $m_{\rm c}$ is unique for all mesons and
provides the $\bar qq$ thresholds after the temperature
exceeds the critical value $\Tc$. The temperature dependence
of the meson properties following from this definition of the model
will be investigated in detail in Sect. \ref{sec:temp}.

\section{Model parameters}

In the present model, there are five parameters: the constituent quark
mass $m$, the 3D UV cut-off parameter $\Lambda$,
the scalar (pseudoscalar) four-quark coupling constant $G_1$,
the vector (axial-vector) four-quark coupling constant $G_2$,  and
the  parameter $\mc$.
To fix our parameters,
we  use  only  four  observables \cite{PartProp}:
the pion weak decay constant
$F_\pi=93$ MeV,
the $\rho\to\pi\pi$ decay constant
$g_\rho^{\rm exp}=6.14$,
the pion mass
$M_\pi=140$ MeV,
the $\rho$-meson mass $M_\rho=770$ MeV,
and the model parameter $\mc$,
determined as $\mc=m(\Tc)$.
The value of $\Tc$ shall be the temperature above which
the lightest meson can decay into free quarks (pion Mott temperature),
which is defined by the formula:
\begin{equation}
\label{Tc}
2m(\Tc)=M_\pi(\Tc).
\end{equation}
We find that $\Tc\approx 186$ MeV, and
$\mc\approx 86$ MeV, see Sect.~6.
To fix $m$, $\Lambda$, $G_1$, and $G_2$,
we use this value of $\mc$ and the following four
equations:
\begin{enumerate}
\item[1)] The Goldberger--Treiman relation
\begin{equation}
\frac{m}{F_{\pi}}=\bar g_{\pi}(M_\pi)=g_\pi(M_\pi)\sqrt{Z}. \label{gold}
\end{equation}
\item[2)] The  $\rho ^0\to \pi^+\pi^-$ decay width. The amplitude
of this process is of the form
\begin{equation}
A_{\rho\to2\pi}=ig_{\rho}^{\rm exp}
(p_{\pi^+}-p_{\pi^-})^\nu \rho^0_{\nu}\pi^+\pi^-.
\end{equation}

In the one-loop approximation, we obtain the
following expression for $g_{\rho}^{\rm exp}$
\begin{eqnarray}
&& g_{\rho}^{\rm exp}=
4Z^{-1}g_{\rho}(M_{\rho})\bar g_{\pi}^2(M_{\pi})
\left[I_2^{(\lambda_{M_\rho},\Lambda)}(M_\rho,m)\right.\nonumber\\
&&\left.\quad+2M_\pi^2 J_{3V}^{\lambda_{M_\rho}}(M_\rho,M_\pi,m)\right],
\end{eqnarray}
where $g_{\rho}(M_\rho)=[\frac23
I_2^{(\lambda_{M_\rho},\Lambda)}(M_{\rho},m)]^{-1/2}$
and the integral $J_{3V}^{\lambda_{M_\rho}}(M_\rho,M_\pi,m)$ is:
\begin{eqnarray}
\label{J3V}
&&J_{3V}^{\lambda_{M_\rho}}(M_\rho,M_\pi,m)
=\quad\frac{1}{M_\rho^2-4M_\pi^2}
\left(I_2^{(\lambda_{M_\rho},\Lambda)}(M_\rho,m)\right.\nonumber\\
&&\left.-\tilde I_2^{(\lambda_{M\rho}\;\Lambda)}(M_\pi,m|M_\rho)
-M_\pi^2 I_3^{\lambda_{M_\rho}}(M_\rho,M_\pi,m)\right),
\end{eqnarray}
where the integrals $I_2^{(\lambda_{M\rho},\Lambda)}(M_\pi,m|M_\rho)$
and $I_3^{\lambda_{M_\rho}}(M_\rho,M_\pi,m)$ are given below:
\begin{eqnarray}
&&\tilde
I_2^{(\lambda_{M_\rho},\Lambda)}(M_\pi,m|M_\rho)=\frac{N_c}{16\pi^2}
\int\limits_{\lambda_{M_\rho}}^{\Lambda}
\frac{{\rm k} d{\rm k} }{E_{\rm k}|\vec{p}|}\nonumber\\
&&\quad\times\ln\left(
\frac{(M_\pi^2+2{\rm k}|\vec{p}|)^2-E_{\rm k}^2M_\rho^2}{(M_\pi^2-2{\rm k}|
\vec{p}|)^2-E_{\rm k}^2M_\rho^2}
\right),\\
&&I_3^{\lambda_{M_\rho}}(M_\rho,M_\pi,m)=
\frac{N_c}{16\pi^2}\int\limits_{\lambda_{M_\rho}}^{\infty}\!
\frac{{\rm k} d{\rm k}}{|\vec{p}|E(M_\rho^2-4E^2)}\nonumber\\
&&\times \left[\ln\left(\frac{(M_\pi^2+2{\rm k}|\vec{p}|)^2
-E^2M_\rho^2}{(M_\pi^2-2{\rm k}|\vec{p}|)^2-E^2M_\rho^2}\right)\right.
\nonumber\\
&&\left.+\frac{M_\rho}{2E}
\ln\left(\frac{M_\pi^4-(E M_\rho-2{\rm k}|\vec{p}|)^2}{M_\pi^4-(E
    M_\rho+2{\rm k}|\vec{p}|)^2}\right)
\right].
\end{eqnarray}
Here, $|\vec{p}|=\sqrt{M_\rho^2/4-M_\pi^2}$ is the 3-momentum of a
pion after the decay of a $\rho$-meson in the rest frame of $\rho$.

The factor $Z^{-1}$ appears due to
$\pi-a_1$ transitions (see \cite{volk86}). From these two
equations one can find $m$ and $\Lambda$.
\item[3)] The coupling constant $G_1$ is determined by the
pion mass formula
\begin{eqnarray}
M_{\pi}^2=
\bar g_{\pi}^2(M_\pi)\left[{1\over G_1}-8 I_1^{\Lambda}(m)\right]~.
\end{eqnarray}
\item[4)] The coupling constant $G_2$  is found from  the  mass
formula for $M_\rho$ \cite{volk86}
\begin{eqnarray}
M_{\rho}^2=\frac{g^2_{\rho}(M_\rho)}{4G_2}=
\frac{3}{8G_2 I_2^{(\lambda_{M_\rho},\Lambda)}(M_\rho,m)}.
\label{rhomass}
\end{eqnarray}
\end{enumerate}
From the gap equation (\ref{gap}), one gets the current quark mass
$m^0$. The results of the  parameter fixing procedure described above are
summarized in Table \ref{tab1}.
For the experimental values of the mass and width of the $\sigma$-meson,
there is a wide uncertainy. The average limits for the mass are
reported to be from 400 MeV to 1200 MeV (see \cite{PartProp,Ishi_96,Svec_92}),
and for the width: from 600 MeV to 1000 MeV. However,
smaller values were also obtained: 290$\pm$54 MeV \cite{Svec97},
119$\pm$13 MeV \cite{Alekseev99}.

\section{The $\sigma$-meson mass and the decay $\sigma \to 2 \pi$}

\begin{figure}[h]
\begin{center}
\includegraphics[scale=0.5]{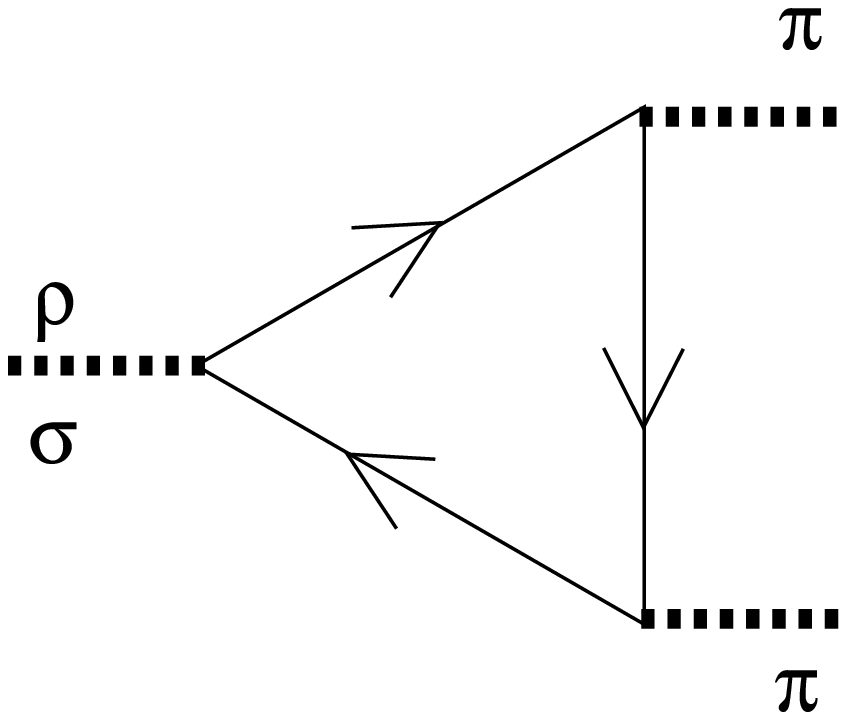}
\end{center}
\caption{}
\label{rs2pi}
\end{figure}

The mass of the $\sigma$-meson is given by  Eq. (\ref{msigma}).
Using this formula for $m_{\rm c}=86$ MeV (see Sect.~6) we obtain
\begin{equation}
M_\sigma=500\;{\rm MeV}.
\label{msig2}
\end{equation}
The decay $\sigma \to 2 \pi$ is described by the quark triangle
diagram (see Fig.~\ref{rs2pi}).

The amplitude of the process  $\sigma \to 2 \pi$ has the form
\begin{eqnarray}
&&A_{\sigma\to2\pi}=8 m g_\sigma(M_\sigma)\bar g_\pi^2(M_\pi)
[I_2^{(\lambda_{M_\sigma},\Lambda)}(M_\sigma,m)\nonumber\\
&&\quad+{\mathcal J}(M_\sigma,M_\pi,m)] \sigma \vec{\pi}^2~~,
\end{eqnarray}
where
\begin{eqnarray}
&&\kern-5mm{\mathcal J}(M_\sigma,M_\pi,m)\!\!=\!-\frac{1}{2}(M_\sigma^2-2M_\pi^2)\!
I_{3}^{\lambda_{M_\sigma}}\!\!(M_\sigma,M_\pi,m).
\end{eqnarray}

Then the decay width of  the  $\sigma$-meson  is
equal to
\begin{eqnarray}
\Gamma_{\sigma\to2\pi}&=&\frac{3}{2\pi}
\left(\frac{m^3(1+\delta)}{g_\sigma(M_\sigma)
 F_\pi^2M_\sigma}\right)^2\sqrt{M_\sigma^2-4M_\pi^2}\nonumber\\
&=& 205~ {\rm MeV}.
\label{sigwidth}
\end{eqnarray}
where
\begin{equation}
\label{deltadef}
\delta=\frac{{\mathcal J}(M_\sigma,M_\pi,m)}
{I_2^{(\lambda_{M_\sigma},\Lambda)}(M_\sigma,m)}
= -0.33 ~.
\end{equation}
Therefore,  one can see that our estimates for the $\sigma$-meson
mass are  in  agreement  with  experimental data
\cite{PartProp} (see also \cite{Ishi_96,Svec_92,Svec97,Alekseev99})
$M_\sigma^{\rm exp} = (400 - 1200)~ {\rm MeV}$.
Let us note that the corrections coming from ${\mathcal J}$
are important for the calculation of the decay width.
Indeed, a similar contribution to the $\rho\to\pi\pi$ decay width
(see (\ref{J3V})) is small, whereas, in the case of the decay
$\sigma\to\pi\pi$, it makes 30\% of the amplitude and decreases
the decay width by half.
From this, one can conclude that the NJL model with the IR cut-off satisfies
both of the low energy theorems together with SBCS and
gives a satisfactory description of  the
low-energy physics of the scalar, pseudoscalar and vector mesons.

\section{Finite temperature case}
\label{sec:temp}

An interesting  application
of our model is the description of meson properties in a hot
and dense medium. The standard NJL model has been already used
for this purpose in  \cite{vogl,hatsuda,munchov}, where the temperature
dependence of quark and   meson masses  and of Yukawa coupling
constants was found.

The calculation of the constituent  quark  and  meson masses at finite
temperature can be done in the imaginary time formalism
\cite{Fetter71,Kapusta89}. In all the quark loop diagrams,
now we sum over Matsubara frequencies $\omega_n=(2n+1)\pi T$ instead
of integrating over the energy component of the internal quark 4-momentum.
As a result,  for the integral $I_1^{\Lambda}(m)$ one has
\begin{eqnarray}
&&I_1^{\Lambda}(m,T)=\nonumber\\
&&\quad-i\frac{N_c}{(2\pi)^4}\int d^4k
\frac{\theta(\Lambda^2-|k_\perp|^2)}{m(T)^2-k^2}
\tanh\left(\frac{E_{k_\perp}}{2T}\right),\\
&&\quad E_{ k_\perp}=\sqrt{|k_\perp|^2+m^2}~.
\end{eqnarray}
The integral $I_2^{(\lambda_M,\Lambda)}(M,m,T)$
for a meson at rest  $P=(M,0,0,0)$ in the rest frame
of the heat bath is given by
\begin{eqnarray}
&&I_2^{(\lambda_M,\Lambda)}(M,m,T)=\nonumber\\
&&\quad-i\frac{N_c}{(2\pi)^4}\int d^4k
{ \theta(\Lambda^2-|k_\perp|^2)
\theta(|k_\perp|^2-\lambda_M^2)\over
  [m(T)^2-k^2][m(T)^2-(k+P)^2] }\nonumber\\
&&\quad\times \tanh\left(\frac{E_{ k_\perp}}{2T}\right)~.
\end{eqnarray}
For the integrals $\tilde I_2$ and $I_3$, one obtains:
\begin{eqnarray}
&&\tilde I_2^{(\lambda_{M_\rho}, \Lambda)}(M_\pi,m,T|M_\rho)=
\frac{N_c}{16\pi^2}\int\limits_{\lambda_{M_\rho}}^{\Lambda}d{\rm k}
\frac{{\rm k} \tanh\left(\frac{E_{\rm k}}{2T}\right) }{E_{\rm k}|\vec{p}|}\nonumber\\
&&\quad\times\ln\left(
\frac{(M_\pi^2-2{\rm k}|\vec{p}|)^2-E_{\rm
    k}^2M_\rho^2}{(M_\pi^2+2{\rm k}|\vec{p}|)^2-E_{\rm k}^2M_\rho^2}
\right),\\
&&I_3^{\lambda_{M_\rho}}(M_\rho,M_\pi,m,T)=
\frac{N_c}{16\pi^2}\int\limits_{\lambda_{M_\rho}}^{\Lambda}\! d{\rm k}
\frac{{\rm k} \tanh\left(\frac{E_{\rm k}}{2T}\right)}{|\vec{p}|E(M_\rho^2-4E^2)}\nonumber\\
&&\times \left[\ln\left(\frac{(M_\pi^2+2{\rm
	k}|\vec{p}|)^2-E^2M_\rho^2}{(M_\pi^2-2{\rm k}|\vec{p}|)^2-E^2M_\rho^2}\right)\right.\nonumber\\
&&\left.+\frac{M_\rho}{2E}
\ln\left(\frac{M_\pi^4-(E M_\rho-2{\rm k}|\vec{p}|)^2}{M_\pi^4-(E
    M_\rho+2{\rm k}|\vec{p}|)^2}\right)
\right].
\end{eqnarray}

\begin{figure}[h]
\begin{center}
\includegraphics[scale=1]{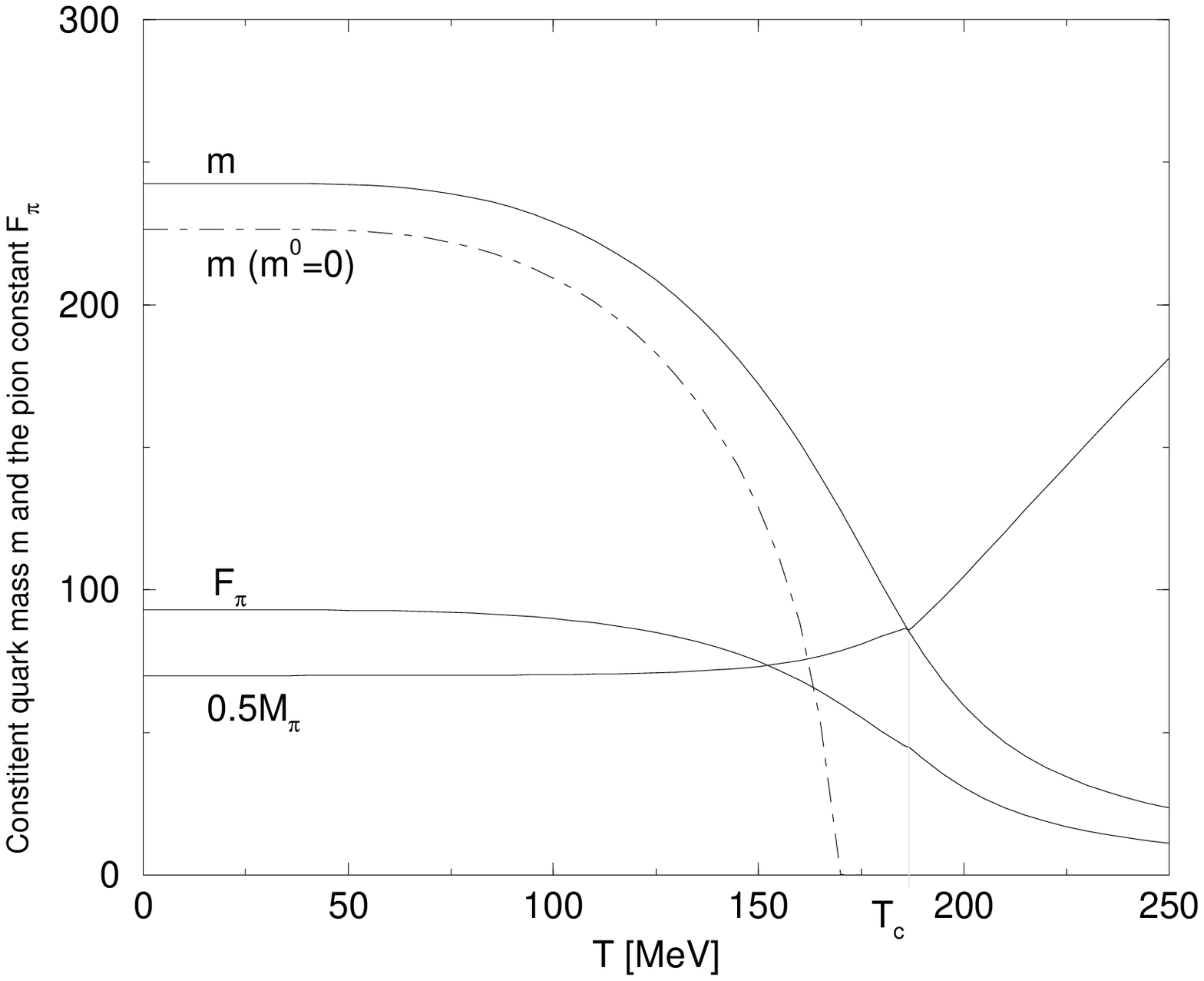}
\end{center}
\caption{}
\label{Qmass}
\end{figure}

\begin{figure}[h]
\begin{center}
\includegraphics[scale=1]{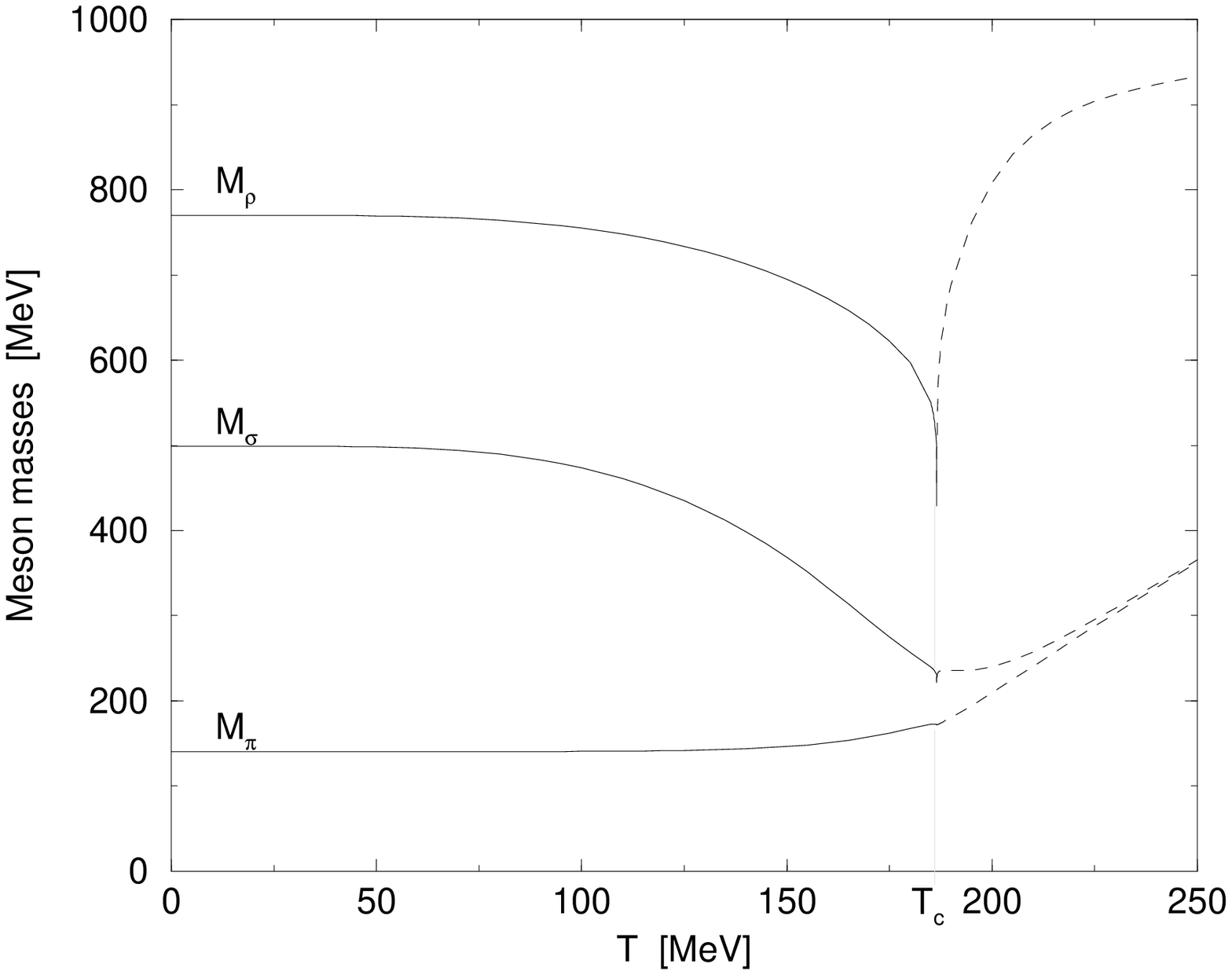}
\end{center}
\caption{}
\label{massT}
\end{figure}

\begin{figure}[h]
\includegraphics[scale=1]{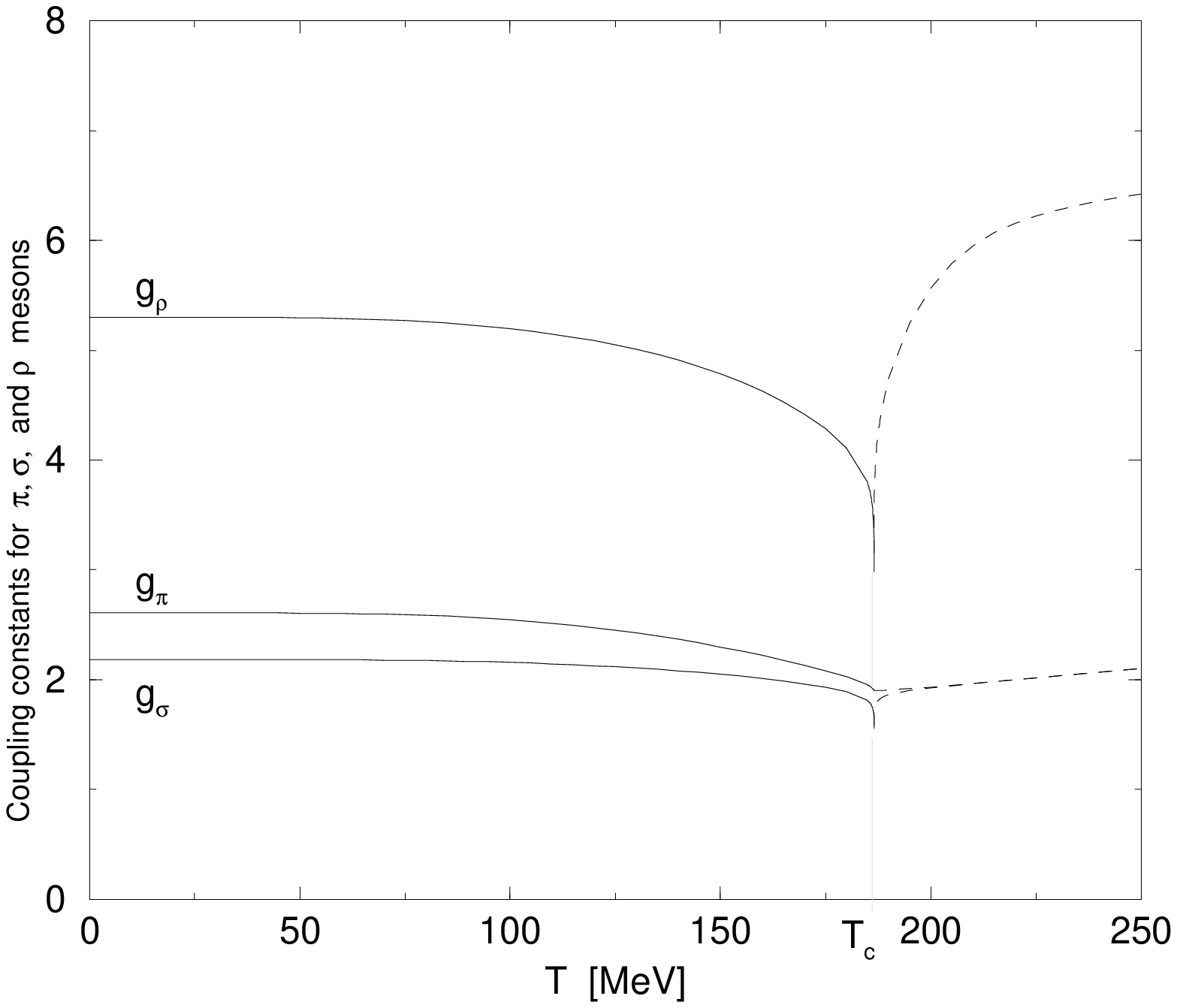}
\caption{}
\label{gT}
\end{figure}

\begin{figure}[h]
\begin{center}
\includegraphics[scale=1]{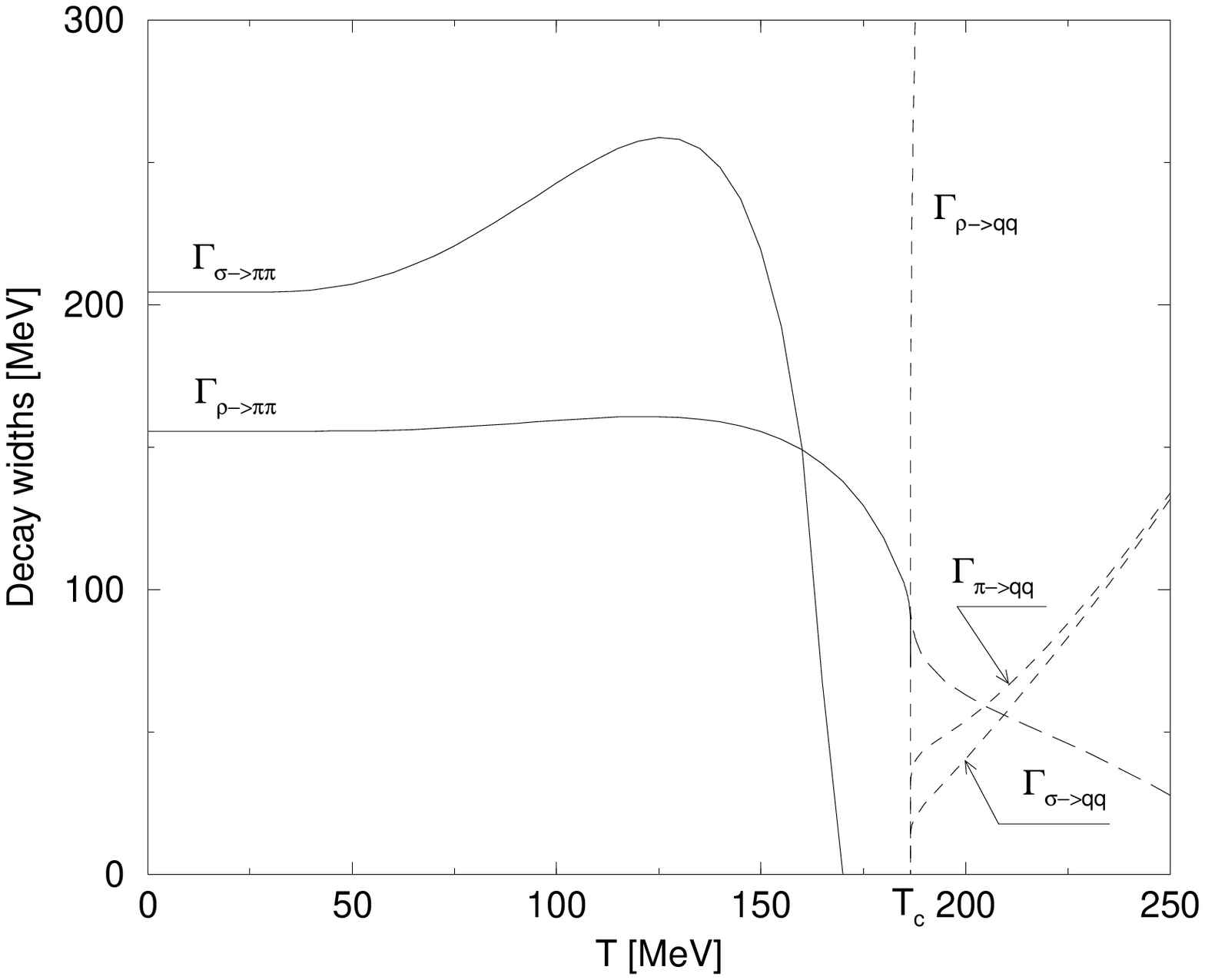}
\end{center}
\caption{}
\label{widths}
\end{figure}

The dependence of the constituent quark mass  on the temperature
is obtained from the gap equation (\ref{gap}) where integral $I_1$ is
already $T$-dependent.
After we know the temperature dependence of
$m$, $I_1$, $I_2$, $\tilde I_2$, and
$I_3$, using formulas (\ref{gpi}), (\ref{pionmass}), (\ref{gsigma}), (\ref{msigma}),
(\ref{gold}), and (\ref{rhomass}),
we can determine the temperature dependence of $F_\pi$,
quark and meson masses,  coupling constants, and  meson decay widths.
The results are shown on
Figs.~\ref{Qmass}-- \ref{widths}.

The $\sigma\to\pi\pi$ decay width is calculated by the formula:
\begin{equation}
\label{sigmadecayT}
\Gamma_{\sigma\to2\pi}(T)=
\frac{3|A_{\sigma\to2\pi}(T)|^2}{32\pi M_\sigma}\sqrt{1-\frac{4M_\pi^2}{M_\sigma^2}}
\coth\left(\frac{M_\sigma}{4T}\right),
\end{equation}
\begin{equation}
A_{\sigma\to2\pi}(T)=\frac{2m(T)Z[1+\delta]\sqrt{I_2^{(\lambda_{M_\sigma},\Lambda)}(M_\sigma,m,T)}}{
I_2^{(\lambda_{M_\pi}\;\Lambda)}(M_\pi,m,T)},
\end{equation}
where $\delta$ is defined in (\ref{deltadef}). The cotangent in (\ref{sigmadecayT})
appeared due to the interaction with the pion gas in the final state.
Analogously, one has for the $\rho$-meson:
\begin{equation}
\Gamma_{\rho\to2\pi}(T)=
\frac{|A_{\rho\to2\pi}(T)|^2M_\rho}{48\pi}\left(1-\frac{4M_\pi^2}{M_\rho^2}\right)^{\frac32}
\coth\left(\frac{M_\rho}{4T}\right),
\end{equation}
\begin{eqnarray}
&&A_{\rho\to2\pi}(T)=\frac{\sqrt{3I_2^{(\lambda_{M_\rho},\Lambda)}(M_\rho,m,T)}}{\sqrt{2}
I_2^{(\lambda_{M_\pi},\Lambda)}(M_\pi,m,T)}\nonumber\\
&&\quad\times\left(1+\frac{2M_\pi^2J_{3V}^{\lambda_{M_\rho}}(M_\rho,M_\pi,m,T)}{I_2^{(\lambda_{M_\rho}\;\Lambda)}(M_\rho,m,T)}\right).
\end{eqnarray}
The quantity  $J_{3V}^{\lambda_{M_\rho}}(M_\rho,M_\pi,m,T)$ is
derived from\sloppy \ $J_{3V}^{\lambda_{M_\rho}}(M_\rho, M_\pi, m)$ (see (\ref{J3V})) by replacing
all integrals with ones depending on temperature.

The critical temperature $\Tc$ is determined by the condition
that the pion mass is equal to the sum of the masses of its constituents
(the Mott point, see Eq.~(\ref{Tc})).
From this, one finds $\Tc\approx 186$ MeV
and $\mc\approx 86$ MeV.
Thus the lightest meson (pion) is also allowed to decay into its
constituents, quarks, at $T\geq \Tc$.
For the IR cut-off scheme considered here,
other mesons also decay into
free quarks if $T\geq \Tc$\footnote{
Note that in the chiral limit ($m^0=0$), chiral symmetry
is restored at $\Tc$.
}.

Finally, we have the following picture.
At low $T$, the  $\sigma$- and $\rho$-mesons
decay mostly into  two pions.
The pion is stable since electroweak decay channels can be neglected here
compared to the strong ones.
At higher $T$, the  $\rho$-meson still has a noticeable decay width.
Unlike the $\rho$-meson,
the width of $\sigma$-meson first rises at $T=100$--150 MeV and then
falls down to zero near $T=170$ MeV.
An increasing of the width is due to the interaction with the pion gas
in the final state, which leads to an additional factor increasing
with temperature.
Above $\Tc$, all mesons are allowed
to decay into quark-antiquark  pairs, and the $\rho$-meson also
decays into pions. It is interesting to note that
the $\sigma$-meson is stable in the temperature range from 170 MeV
to $\Tc$. Here, only its electroweak decays are possible, they are
small and can be neglected. Thus we have obtained almost stable scalar
meson states freely propagating through hot matter as a precursor of the
chiral/ deconfinement transition.

To estimate the decay widths of $\pi$-, $\sigma$- and $\rho$-mesons into free
quarks, one should evaluate the imaginary part of the corresponding meson
propagator that appears if $T>\Tc$. One thereby has:
\begin{eqnarray}
\Gamma_{\pi\to\bar qq}(T)&=&\frac{M_\pi\mathrm{Im}
  I_2^{(\lambda_{M_\pi},\Lambda)}(M_\pi,m,T)}{
\mathrm{Re} I_2^{(\lambda_{M_\pi},\Lambda)}(M_\pi,m,T)},\\
\Gamma_{\sigma\to\bar qq}(T)&=&\frac{[M_\sigma^2-4m(T)^2]\mathrm{Im}
  I_2^{(\lambda_{M_\sigma},\Lambda)}(M_\sigma,m,T)}{M_\sigma
\mathrm{Re} I_2^{(\lambda_{M_\sigma},\Lambda)}(M_\sigma,m,T)},\\
\Gamma_{\rho\to\bar qq}(T)&=&\frac{M_\rho\mathrm{Im}
  I_2^{(\lambda_{M_\rho},\Lambda)}(M_\rho,m,T)}{
\mathrm{Re} I_2^{(\lambda_{M_\rho},\Lambda)}(M_\rho,m,T)}.
\end{eqnarray}
The results are shown in Fig.~\ref{widths} by dashed and dash-dotted lines.

\section{Discussion and conclusion}

In this paper, we have investigated an extension of the NJL model for the
light nonstrange meson sector of QCD, where the interaction of
$u$- and $d$-quarks is represented by four-fermion vertices and the
phenomenon of quark confinement is taken into account through the
elimination  of non-physical quark-antiquark thresholds. This extension
of the NJL model  describes  properties of the  $\pi$-, $\sigma$- and
$\rho$-mesons in  satisfactory agreement with  experiment and with  low-energy
theorems.
The model parameters are obtained by fitting the model so that it reproduces
the experimental values of the pion and $\rho$-meson masses, the pion decay
constant $F_\pi$, and the $\rho$-meson decay constant $g_\rho$.
Moreover, it has been shown that, for the $\pi$-, $\sigma$- and $\rho$-mesons,
unphysical quark-antiquark thresholds do not appear up to the critical
temperature if the IR cut-off
of the form introduced here  is applied.

Let us emphasize that
in our model, unlike the standard NJL model \cite{eguchi,ev83,volk86,ebert86},
we have two cut-offs: the UV cut-off that eliminates the UV divergences
and the IR cut-off which provides the  confinement of quarks.
The UV cut-off  determines the dimension of the domain
of SBCS where quarks are bosonized.
It is chosen to be the same for all sorts of mesons.
The second cut-off, $\lambda_M$, is introduced into the model in
analogy with the bag-model \cite{bag,bag1,bag2} and describes
finite dimensions
of mesons. We suppose that
heavier mesons have smaller radii, therefore  the IR cut-off is
chosen different for various mesons,
being roughly  proportional to the meson mass.
On the other hand, from  the requirement of the absence of quark-antiquark
thresholds for $T<\Tc$, we determine more  certainly the  form of the IR cut-off.
It is easy to make sure that the IR cut-off of the form (\ref{cutoff})
satisfies this condition.

The critical temperature is defined as the one at which
the pion mass equals the sum of the masses of constituent quarks
(the so called Mott point). Thus, after the matter reachs $\Tc$,
the pion became unstable as the other mesons and
they all are allowed to decay into free quarks under
such conditions. This scenario is provided by the IR cut-off
scheme implemented in the present work.
Although, only scalar, pseudoscalar, and vector mesons
have been considered, the axial-vector meson can also be
treated the same way, and no unphysical $\bar qq$ thresholds
will appear for it if $T<\Tc$.

Let us note that the introduction of the IR cut-off has not
dramatically changed the basic model parameters when compared to a
the standard NJL model case \cite{munchov} with $\lambda_P=0$.
For example, the UV cut-off increased from 1.03 GeV up to  1.09 GeV,
the constituent quark mass decreased from 280 MeV  down to 242 MeV,
the constant $G_1$ also decreased from 3.48 GeV$^{-2}$ down to 2.98 GeV$^{-2}$.
The current quark mass has almost not changed  its value 2.1 MeV.
The mass and width of the $\sigma$-meson, obtained in this model,
are in their experimental bounds \cite{PartProp,Svec97}.

The mechanism of the confinement of quarks that we introduced in our
model allows us to take into account the dependence of various
quantities on  external momenta.
As a result, we managed to estimate additional contributions
to the amplitudes of $\sigma\to\pi\pi$, $\rho\to\pi\pi$
that are proportional to meson masses squared.
It turned out that for the decay $\rho\to\pi\pi$
these corrections were small, and for the decay $\sigma\to\pi\pi$
they made about 30\% of the amplitude and
decreased the decay rate by half.
Taking into account these contributions may
be  important if one wants to describe such quantities as
form factors occurring in various processes, meson radii,
scattering lengths, polarizability etc.

The $\sigma$- and $\rho$-mesons, considered here,
can play an important r\^ole as intermediate  resonances
in the processes occurring in the hot hadron matter
created in ultra-rela\-ti\-vis\-tic heavy-ion collisions.
In particular, a proper consideration of these states  can be useful for
the explanation and
prediction of signals  witnessing CSR and
the quark deconfinement  at the transition of the hadron matter  to the
quark-gluon plasma phase and vice versa;
for example, the low-mass
dilepton enhancement  observed by the CERES collaboration
\cite{CERES,Rapp}.

All calculations in our model are performed in the Hartree-Fock approximation
which does not take into account the next to $1/N_c$ contributions.
For applications to the situation in heavy-ion collisions, where a hot and
dense fireball of mesons (predominantly pions) is formed, it is
of interest to calculate contributions coming from intermediate
pion resonances in the loop diagrams. The properties of the
$\rho$-meson  can also be modified due to the cloud of
pions in  hot matter \cite{Wambach}.
An analogous situation is expected for the $\sigma$-meson \cite{Oertel}.
In our future work we suppose to investigate the next to leading order
corrections in the $1/N_c$ expansion.

\section*{Acknowledgement}
DB thanks Sebastian Schmidt for useful comments and discussions.
This work has been supported by RFBR Grant No 00-02-17190 and the
Heisenberg-Landau program, 2001. GB, MKV and VLYu acknowledge support
by the Ministery for Education and Research of Mecklenburg-Vorpommern and
by the DFG Graduiertenkolleg {\it Strongly correlated many-particle systems}.

\clearpage

\section{Table captions}
\begin{enumerate}
\item Model paramters $\mc, m, \Lambda, G_1, G_2$ and the mass and
width of the $\sigma$-meson.
\end{enumerate}

\section{Figure captions}
\begin{enumerate}
\item Three domains for quark propagation in coordinate space and their
relation to  the UV and IR cut-offs in the momentum space.
\item The quark-loop diagram for the polarization
operator of $\sigma$ and $\pi$.
\item The triangle quark diagram describing the decay of
$\rho$- and $\sigma$-mesons into two pions.
\item The dynamical quark mass: physical and  in the chiral limit
  ($m^0=0$, dash-dotted line), and the pion weak coupling constant $F_\pi$,
  and a half of the pion mass.
\item Masses of $\sigma$, $\pi$, and $\rho$.
The region above $\Tc$ is represented by dashed lines.
\item The coupling constants $g_\pi$, $g_\sigma$, and $g_\rho$
 as functions of temperature. The region
  above $\Tc$ is represented by dashed lines.
\item The decay widths of $\pi$,  $\sigma$ and $\rho$ ($\pi\pi$ and $\bar
  qq$ channels). The region
  above $\Tc$ is represented by dashed lines.
\end{enumerate}

\clearpage
\section{Tables}

\begin{table}[hbt]
\caption{}
\label{tab1}
\begin{center}
\begin{tabular}{|l|l|}
\hline
$m_{\rm c}$ { [MeV] } & 86 \\
$m$  [MeV] &  242 \\
$m^0$ [MeV] &  2.1 \\
$\Lambda$  [GeV] & 1.09 \\
$G_1$  [GeV]$^{-2}$ &  2.98\\
$G_2$  [GeV]$^{-2}$&  11.8\\
$M_\sigma$  [MeV] & 500\\
$\Gamma_\sigma$  [MeV] & 205 \\
\hline
\end{tabular}
\end{center}
\end{table}

\clearpage
\section{Figures}



\end{document}